\documentclass[aps,floatfix,prb,twocolumn,superscriptaddress]{revtex4-2}
\pdfoutput=1
\usepackage[english]{babel}
\usepackage{amsmath}
\usepackage{graphicx}
\usepackage{float}
\usepackage{bm}
\usepackage{multirow}
\usepackage{dcolumn}
\usepackage{color}
\newcommand{\B}{\textcolor{blue}} 
\usepackage{hyperref}
\hypersetup{colorlinks,linkcolor={blue},citecolor={blue},urlcolor={blue}}
\usepackage{siunitx}

\begin{document}

\title{Infrared ellipsometry study of the charge dynamics in K$_3$\,\!p-terphenyl}

\author{Qi He}
\email[]{qi.he@unifr.ch} 
\affiliation{Department of Physics, University of Fribourg, 1700 Fribourg, Switzerland}

\author{P. Marsik}

\author{F. Le Mardel\'e}

\author{B. Xu}

\affiliation{Department of Physics, University of Fribourg, 1700 Fribourg, Switzerland}

\author{M. Sharma}
\affiliation{School of Science and Technology, Physics Division, University of Camerino, 62032 Camerino, Italy}

\author{N. Pinto}
\affiliation{School of Science and Technology, Physics Division, University of Camerino, 62032 Camerino, Italy}
\affiliation{Advanced Materials Metrology and Life Science Division, INRiM, 10135 Torino, Italy}

\author{A. Perali}
\affiliation{School of Pharmacy, Physics Unit, University of Camerino, 62032 Camerino, Italy}

\author{C. Di Nicola}
\affiliation{School of Science and Technology, Chemistry Division, University of Camerino, 62032 Camerino, Italy}

\author{C. Pettinari}
\affiliation{School of Science and Technology, Chemistry Division, University of Camerino, 62032 Camerino, Italy}

\author{D. Baeriswyl}
\affiliation{Department of Physics, University of Fribourg, 1700 Fribourg, Switzerland}

\author{C. Bernhard}
\email[]{christian.bernhard@unifr.ch} 
\affiliation{Department of Physics, University of Fribourg, 1700 Fribourg, Switzerland}

\date{\today}
\begin{abstract}
We report an infrared ellipsometry study of the charge carrier dynamics in polycrystalline K$_{x}$\,\!p-terphenyl samples with nominal $x=3$, for which signatures of high-temperature superconductivity were previously reported. The infrared spectra are dominated by two Lorentzian bands with maxima around 4\,000\,cm$^{-1}$  and  12\,000\,cm$^{-1}$ which, from a comparison with calculations based on a H\"uckel model are assigned to intra-molecular excitations of $\pi$ electrons of the anionic p-terphenyl molecules. The inter-molecular electronic excitations are much weaker and give rise to a Drude peak and a similarly weak Lorentzian band around  220\,cm$^{-1}$. A dc resistivity of about 0.3 $\Omega$cm at {300}{K} is deduced from the IR data, comparable to values measured by electrical resistivity on a twin sample. 
The analysis of the temperature dependence of the low-frequency response reveals a gradual decrease of the plasma frequency and the scattering rate of the Drude peak below {300}{K} that gets anomalously enhanced below {90}{K}. The corresponding missing spectral weight of the Drude peak appears blue-shifted towards the Lorentz-band at 220\,cm$^{-1}$. This characteristic blue-shift signifies an enhanced localization of the charge carriers at low temperatures and contrasts the behavior expected for a bulk superconducting state for which the missing spectral weight would be redshifted to a delta-function at zero frequency that accounts for the loss-free response of the superconducting condensate. Our data might still be compatible with a filamentary superconducting state with a volume fraction well below the percolation limit for which the spatial confinement of the condensate can result in a plasmonic resonance at finite frequency.
\end{abstract}
\maketitle

%
\section{Introduction}
Organic molecular solids have long been understood as insulators. 
This changed in the Seventies and early Eighties with the synthesis of conducting charge-transfer salts, such as TTF TCNQ~\cite{Ferraris_73, Coleman_73, Torrance_79} and (TMTSF)$_2$PF$_6$~\cite{Bechgaard_80}. In these materials, reviewed in Ref. \cite{Jerome_04}, the relatively strong overlap between electronic wave functions due to the face-to-face packing of molecules leads to one-dimensional bands close to the Fermi energy along the stacks. The Bechgaard salts (TMTSF)$_2$X, where X=PF$_6$, ClO$_4$ etc., show a variety of phases, including spin-density waves, charge order and, in particular, superconductivity~\cite{Jerome_80, Parkin_81}. 
The possibility of superconductivity in a one-dimensional organic metal, possibly even up to high temperatures, had been theoretically proposed already in 1964 by W. A. Little~\cite{Little_64}. Such an organic high-T$_c$ superconductor would enable a wide range of promising applications for biological devices, photovoltaics (solar cells), memory materials, optoelectronic devices and other advanced technologies. 

The critical temperatures of superconducting Bechgaard salts are low, of the order of {1}{K}. Higher $T_c$'s were reached in the quasi-two-dimensional organic compounds (BEDT-TTF)$_{2}$X~\cite{Williams_91}, up to {14.2} {K}~\cite{Taniguchi_03}. These materials exhibit a variety of exotic phases, from spin liquids to unconventional superconductivity~\cite{Powell_11, Ardavan_12, Dressel_20}.

The discovery of C$_{60}$ (sometimes simply called fullerene) in 1985 marked the beginning of a new era in the research on organic molecular compounds~\cite{Forro_01}. Solid C$_{60}$ is a cubic material and therefore quite different from the quasi-one- and quasi-two-dimensional charge transfer salts mentioned above. The inter-molecular overlap is very small in all directions (and not only between chains as in TTF TCNQ or between planes as in graphite) and therefore the molecular orbitals are ``good starting points for describing electronic bands in solid C$_{60}$''~\cite{Dresselhaus_96}. In this sense the fullerenes are quasi-zero-dimensional. Solid C$_{60}$ is semiconducting, with an optical gap of about 2 eV~\cite{Degiorgi_98}. Doping with potassium renders the material metallic and even superconducting, with $T_c$ = {18}{K} for K$_3$C$_{60}$~\cite{Hebard_91}. At present, the commonly accepted record value for an organic superconductor of $T_c$ = {38}{K} is held by Cs$_{3}$C$_{60}$ under high pressure~\cite{Takabayashi_09}. It seems to be widely accepted that the three-fold degeneracy of the lowest unoccupied molecular orbital (LUMO) plays an important role in the superconductivity of alkali-metal-doped fullerene, through the enhancement of the electron-phonon coupling by the Jahn-Teller effect~\cite{Gunnarsson_97}, but electron correlation effects clearly are also relevant, especially for Cs$_3$C$_{60}$, a Mott insulator under ambient pressure~\cite{Capone_09}.

In 2010, superconductivity has been reported for the first time in a polycyclic aromatic hydrocarbon, namely picene (C$_{22}$H$_{14}$) doped with potassium~\cite{Mitsuhashi_10, Kubozono_16}, with a surprisingly large critical temperature, $T_c=18$ K. Since then, other materials of the same type have been found to become superconducting by doping, notably phenanthrene (C$_{14}$H$_{10}$) with  $T_c=5$ K~\cite{Wang_11}, dibenzopentacene (C$_{30}$H$_{18}$) with $T_c=33$ K~\cite{Xue_12}, and  pentacene (C$_{22}$H$_{14}$) with $T_c=4.5$ K~\cite{Nakagawa_16}. The interpretation of some of these experiments has been questioned~\cite{Heguri_15}. In fact, there are problems with hydrocarbon superconductors~\cite{Kubozono_16}, for instance the shielding fraction obtained from the magnetization is very low, typically of the order of 1\%.\\
More recently, signatures of even higher $T_c$ values of {43}{K} or even up to {123}{K} have been reported for p-terphenyl (C$_{18}$H$_{14}$) doped with potassium, with a nominal K-content of $x=3$~\cite{Wang_17b, Wang_17c,Neha_18}. Evidence for a superconducting transition has been reported here \B{mainly} from magnetic susceptibility measurements. The analysis of the magnetic susceptibility data, however, yields a small superconducting volume fraction on the order of only a few percent. This latter result has been confirmed by a magnetization study of an independent group on polycrystalline K$_3$\,\!p-terphenyl samples that were synthesized under high pressure conditions~\cite{Liu_17}. Meanwhile, gap-like features in the electronic excitation spectrum that are reminiscent of a superconducting order have been observed with angle-resolved photoemission spectroscopy (ARPES)~\cite{Li_19} and with scanning tunneling spectroscopy~\cite{Ren_19} on the surface of p-terphenyl crystals on which monolayers of potassium have been evaporated under ultrahigh vacuum condition. Signatures of a possible high-$T_c$ superconducting phase have also been reported by electrical measurements~\cite{pinto2020potassium}. 
The nominal composition of the bulk samples for which signatures of high $T_c$ superconductivity have been reported amounts to K$_{3}$\,\!p-terphenyl~\cite{Wang_17c}. Nevertheless, it remains disputed whether in these samples the potassium is homogeneously distributed. Accordingly, the superconducting phase might have a different K-content or even originate from clusters of K-rich material for which the composition is presently unknown as discussed, e.g., in Refs.~\cite{Zhong_18, Yan_18, Guijarro_20}.
Moreover, the high sensitivity of the K$_x$\,\!p-terphenyl samples to oxygen and moisture causes them to decompose rather rapidly under ambient conditions. This makes it quite difficult to study, for example with infrared spectroscopy, their bulk-like free carrier response and the signatures of a possible superconducting state.\\
The infrared spectroscopy technique probes the complex dielectric function of a material and thus provides valuable information about the dynamics of the mobile or weakly bound charge carriers as well as about their interband transitions and the related band structure ~\cite{Wooten_73, Dressel_02}. In organic conductors the low-energy electronic response is typically governed by the inter-molecular excitations for which the magnitude depends on the stacking and the bonding between the molecules and thus can be strongly anisotropic and exhibit large variations between different materials. The response at higher energy and the band gaps are typically characteristic of the intra-molecular excitations of the individual molecules and thus can serve as \textquotedblleft fingerprints" of their structure and their ionization level. Prominent examples of materials for which these electronic properties have been extensively studied with infrared spectroscopy are the Bechgaard salts, like (TMTSF)$_{2}$AsF$_{6}$~\cite{Ng_85} and k-(BEDT-TTF)$_{2}$X~\cite{Ito_04, Eldridge_91, Kornelsen_89, Kornelsen_91, Dressel_12} or the doped fullerenes~\cite{Degiorgi_98}.\\
Especially powerful is the spectroscopic ellipsometry technique which measures the change of the polarization state of the light upon reflection from the sample surface, rather than the intensity. It provides direct access to the real and the imaginary parts of the dielectric function of a given material~\cite{Aspnes_83, Korte_98}. Unlike the conventional reflection technique, it does not require reference measurements which are typically done by replacing the sample with a reference mirror or by gold coating the sample surface, nor a Kramers-Kronig transformation, for which the measured data have to be extrapolated toward zero and infinite frequency. The ellipsometry techniques is therefore well suited to measure the dielectric function of samples that have to remain under inert gas atmosphere or very high vacuum condition, before and during the measurements, and therefore are sealed in a closed cell with suitable windows for the optical access for which it is difficult to perform accurate reference measurements.  
Ellipsometry is commonly applied in the near-infrared, visible and UV ranges, where ellipsometers are commercially available. Meanwhile, some groups, including ours, have built ellipsometers that operate in the far-infrared (FIR)~\cite{Schubert_04, Bernhard_04} and the terahertz (THz) ranges~\cite{Matsumoto_11, Morris_12, Marsik_16} and used them to study the electronic properties of various insulators, semiconductors, metals and even superconductors, including organic materials~\cite{Hinrichs_05}.\\ 
In the following we present an infrared ellipsometry study of a polycrystalline sample of heavily K-doped p-terphenyl that decompose rapidly under ambient condition and thus need to be handled and measured under dry inert gas atmosphere or under high vacuum. For this purpose we have equipped our infrared ellipsometer with an optical cell which has windows made from undoped silicon (or alternatively from KBr) that provide optical access to the sample and enable ellipsometry measurements.
In detail, in section II, we provide the technical details about the sample preparation and the ellipsometry and electronic transport experiments. In Section III, we discuss the infrared data and analyze them in terms of a simple model, with a Drude part and several Lorentz oscillators. In Section IV, the main absorption peaks are attributed to intra-molecular excitations and interpreted semi-quantitatively, according to H\"uckel theory. Inter-molecular excitations are made responsible for the weak low-frequency absorption. Section V presents the temperature dependence of both the low-frequency optical conductivity and the electrical resistivity. The observed spectral weight loss of the Drude peak is consistent with the upturn of resistivity at low temperatures. Our main findings and conclusions are summarized in Section VI. Finally, Appendix~\ref{sec:huckel} presents details about orbitals and optical transitions of the p-terphenyl molecule, within H\"uckel's tight-binding approximation.

\par 

\section{Experimental procedure}

Polycrystalline potassium doped p-terphenyl, K$_{x}$\,\!p-terphenyl (KPT), was synthetised at the University of Camerino, as described in Ref.~\citep{pinto2020potassium}. Inside a glove box, the potassium was cut in little pieces and mixed with the p-terphenyl in a molar ratio of 3:1 as to achieve a nominal composition of $x=3$. To enable the chemical reaction, the mixture was sealed in a glass tube under argon atmosphere and heated to a temperature of {503}{K} at which it was kept for 2 hours. The resulting black powder, very sensitive to oxygen and water, was kept under dry and inert atmosphere of pure Argon gas.\\
Temperature dependent electrical resistivity and current-voltage {\it{I-V}} characteristics were carried out on compressed KPT powder. A novel, redesigned sample holder, with respect to those used in Ref.~\cite{pinto2020potassium}, was fabricated and used in this work. The sample holder consists in a Teflon cylinder of 12 mm of internal diameter, closed at the two ends by copper electrodes that are electrically insulated against each other. The cylinder was filled with 40 mg of KPT powder, under a dry Ar atmosphere, and pressure was applied to the two electrodes and kept during the measurement. For this study, the powder was compressed at about 0.45 MPa, in all investigated samples. Resistivity and {\it{I-V}} characteristics were measured with a pulsed technique, sourcing short current pulses of 1.1 ms of duration, either fixed or with a varying intensity, by a source-meter Keysight mod. B2912A, in the 4-wire geometry of contacts. Compared to the standard dc current technique, the pulsed one allows to minimize thermal electromotive force offsets~\cite{Keithley} and to source higher current intensities. No appreciable differences were detected in the electrical transport properties measured by the pulsed and constant dc current techniques, at low current density.
The sample temperature has been detected with a calibrated Si diode thermometer (Lakeshore mod. DT-670).\\
For the optical measurements pellets with a diameter of {12}{mm} and a thickness of about {1}{mm} were pressed from the K$_{3}$\,\!p-terphenyl powder. Such pellets have been prepared from three different growth batches which all have the same nominal composition of K$_{3}$\,\!p-terphenyl. For sending the samples to the University of Fribourg, they were enclosed in sealed plastic containers filled with Argon gas.

\begin{figure}[htb] 
	\includegraphics[width=0.85\linewidth]{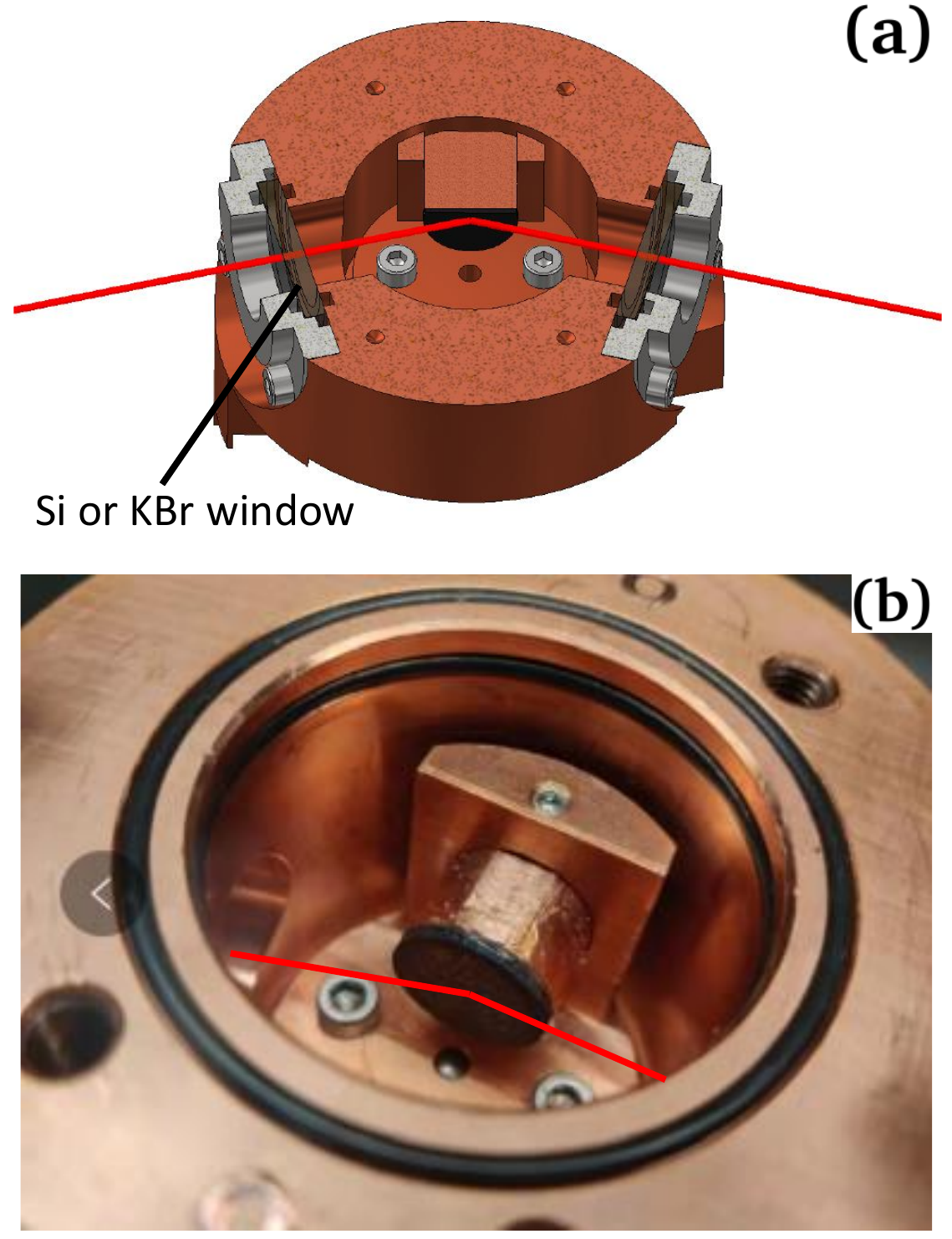}
	\caption{(a) Schematic drawing of the cell for infrared ellipsometry measurements under dry and inert gas atmosphere. Red lines show the beam path of the incident and the reflected photons. (b) Photographic image of the inside of the cell with a K$_{3}$\,\!p-terphenyl sample (black pellet) mounted. }
	\label{fig1}
\end{figure}

In Fribourg the containers were transferred to a glove-box filled with dry Ar gas where they were opened and the samples were mounted in a home made optical cell. A schematic drawing and an image of this optical cell with a sample (black pellet) mounted are displayed in Fig.~\ref{fig1}. The sample was glued to a copper holder with vacuum grease to enable a good thermal contact. The cell was sealed inside the glove box under dry Ar gas atmosphere using viton o-rings. The optical access for the  ellipsometry measurements was enabled by two windows from pure silicon (sealed with viton rings) that are transparent in the entire infrared range up to 8\,000~cm$^{-1}$, except for a narrow region between 500 and 600 cm$^{-1}$ with strong multiphonon absorption. Alternatively, we used KBr windows that are transparent from 350 cm$^{-1}$ up to the UV range. Subsequently, the sealed optical cell with the sample inside was mounted on the cold head of a He-flow cryostat (from Cryovac) that enables temperatures ranging from {6}{K} to {400}{K}. Our home built infrared ellipsometer setup is attached to a Fast Fourier Transform Infrared (FTIR) spetrometer (Bruker VERTEX 70\,v). Its outline is described in Ref.~\cite{Bernhard_04}. The ellipsometric spectra have been recorded at an angle of incidence of $\phi$ = 75$^{\circ}$ for the temperature range from {300}{K} to {6}{K}. In the far-infrared from about 50 to 700~cm$^{-1}$ we used a rotating analyser setup with polarizers made from wiregrids that are evaporated on thin polyethylene foils. The mid-infrared range from about 600 to 5\,000~cm$^{-1}$ was measured with a rotating compensator setup based on a ZnSe prism and wire grid polarizers on KRS-5 substrates. For the near-infrared range from 5\,000 to 10\,000~cm$^{-1}$ we used a rotating analyser configuration with Glan-Thompson polarizers from Calcite. For the entire spectral range a He-cooled bolometer was used as detector.  

\section{Analysis of the room-temperature spectra}\label{sec:spectra}

\begin{figure}[htb]  
	\includegraphics[width=0.95\linewidth]{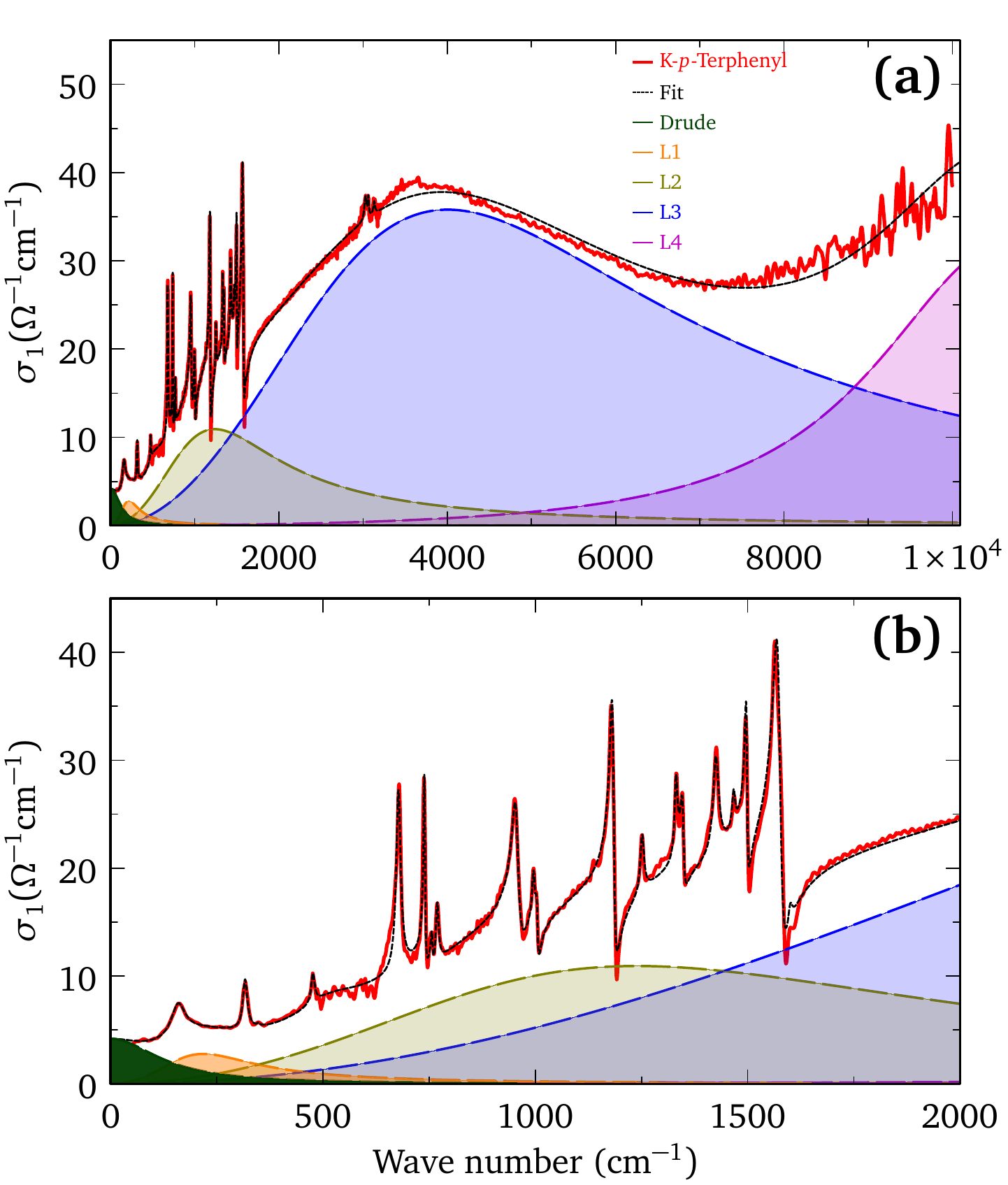}
	\caption{(a) Broad-band spectrum of the optical conductivity at {300}{K} of polycrystalline K$_{3}$\,\!p-terphenyl. Solid lines and corresponding shaded areas show the contributions of the Drude peak (dark green) and of the broad Lorentz oscillators, as obtained with the fit function (\ref{eq1}). (b) Magnified view of the low-frequency part of the optical conductivity spectrum.}
	\label{fig2}
\end{figure}

Figure~\ref{fig2}(a) displays the spectrum  at {300}{K} of the optical conductivity of a polycrystalline K$_{3}$\,\!p-terphenyl sample for the frequency range from about 50 to 10\,000~cm$^{-1}$ (6~meV -- 1.25~eV). Figure~\ref{fig2}(b) shows a magnified view of the low-frequency part of the spectrum up to 2\,000~cm$^{-1}$. Similar spectra have been obtained by samples belonging to three different growth batches of K$_{3}$\! p-terphenyl. 

The electronic part of the spectrum of the complex dielectric function $\varepsilon(\omega)$ has been fitted to a Drude peak and four Lorentz oscillators,
\begin{align}\label{eq1}
   \varepsilon(\omega) = \varepsilon_\infty -\dfrac{\omega_{p0}^{2}}{\omega^{2}+i\omega\Gamma_{0}}+\sum^4_{i=1} \frac{\omega_{pi}^2}{{\omega_{i}^2-\omega^2-i\omega\Gamma_{i}}},
\end{align}
where $\omega_{p0}$ and $\Gamma_{0}$ are the plasma frequency and the scattering rate of the Drude model, respectively, $\omega_{i}$ is the eigenfrequency, $\Gamma_{i}$ the linewidth and  $\omega_{pi}$ the plasma frequency of the $i$th Lorentz oscillator and $\varepsilon_\infty$ is the high frequency dielectric constant representing interband transitions well above the measured range. The individual contributions of the Drude and Lorentz oscillators are shown by solid lines and corresponding shadings. The parameters obtained from the fit are listed in Table~\ref{TableLat}, which also shows the spectral weights of the Drude and Lorentz bands, defined as

\begin{equation}\label{eq:sum_2}
\text{SW}_i  \equiv \int_{0}^{\infty} \sigma_{1i}(\omega) \,d\omega = \dfrac{\pi}{2}\varepsilon_0\omega_{pi}^{2} = \dfrac{\pi N_i e^{2}}{2m_i}.
\end{equation}
Here $e$ is the electron charge, while $m_i$ and $N_i$ are, respectively, the effective mass and the density of electrons involved in the $i$th component of the spectrum. 

\begin{table} 
\centering
$
\begin{array}{crrrr}
\mbox{Transition}& \omega_i & \Gamma_i  & \omega_{pi} & \mbox{SW}_{i} \\ 
\hline
\mbox{Drude}&--&311& 247 & 1586\\
L_1&218 &285& 325 & 2746\\
L_2 &1239& 1793&1084  & 30551\\
L_3 & 4003&6199& 3647  & 345815\\
L_4 & 10700   &4015& 2786 &  201807\\
\end{array}
$
\caption{Parameters of the Drude and Lorentz oscillators derived by fitting the infrared spectra of K$_{3}$\,\!p-terphenyl with Eq. (\ref{eq1}). Units are: cm$^{-1}$ for $\omega_i$, $\omega_{pi}$, $\Gamma_i$; $\Omega^{-1}${cm}$^{-2}$ for the spectral weights (SW)$_i$.}
\label{TableLat}
\end{table}

\begin{figure}[htbp] 
	\includegraphics[width=0.95\linewidth]{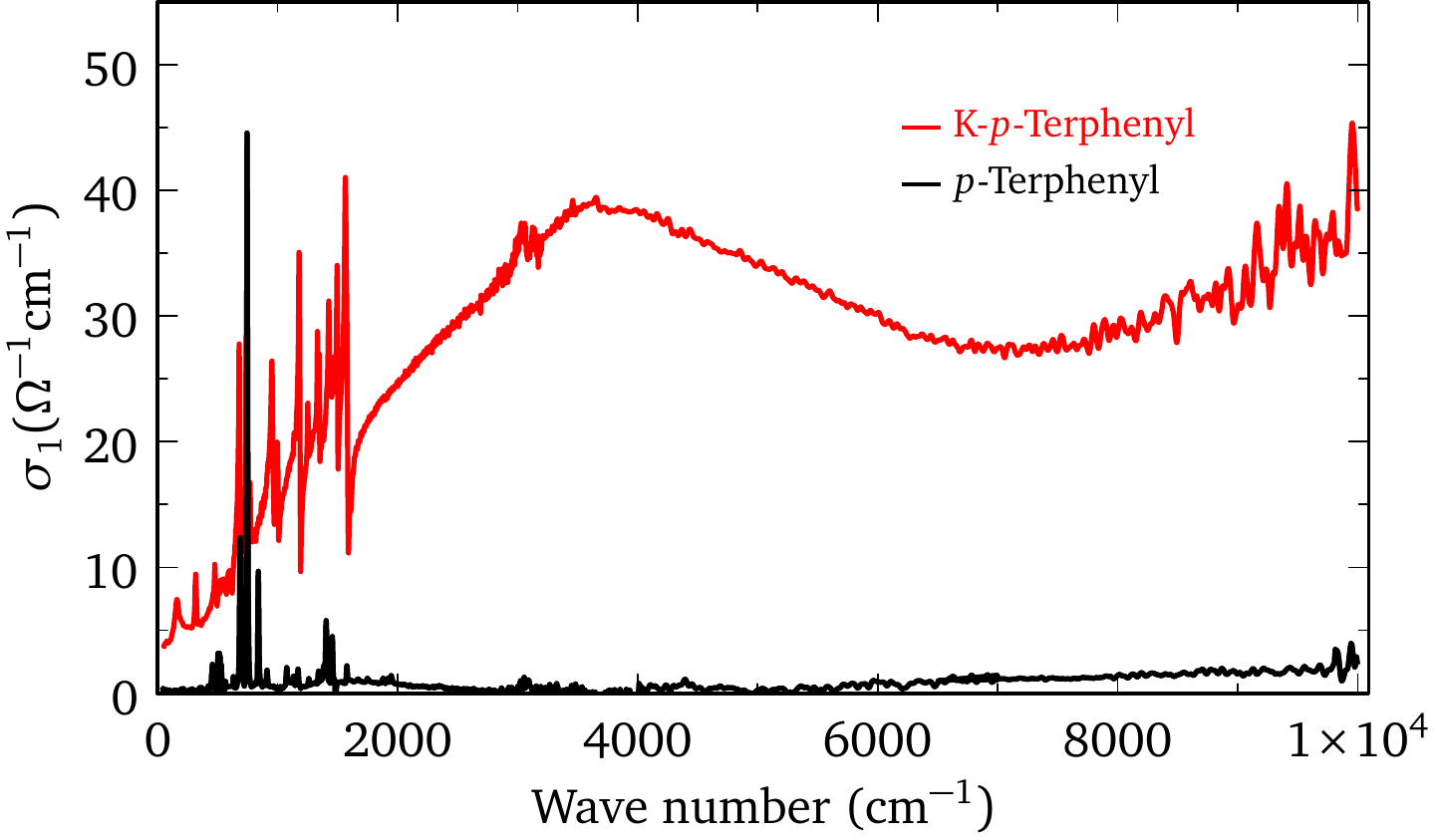}
	\caption{Comparison of the optical conductivity spectra of polycrystalline samples of pristine p-terphenyl (black) and K$_{3}$\,\!p-terphenyl (red) at {300}{K}. }
	\label{fig8}
\end{figure}

Unfortunately, we could not extend the ellipsometry measurements to the range above 10\,000~cm$^{-1}$, to determine more accurately the center frequency and the spectral weight of the $L_4$ band. This is due to scattering and depolarisation effects which appear if the wavelength of the photons reaches the grain size of the polycrystalline samples. Another problem is related to luminescence effects which become prominent as the photon energy approaches the band gap~\cite{Kumar_79, Andrushenko_01}. Nevertheless, since ellipsometry provides independent measures of the real and imaginary parts of the dielectric function (or related optical response functions), the data at hand allow us to obtain at least a rough estimate of the center frequency and the spectral weight of the $L_4$ band. 

The infrared spectrum also contains several much narrower peaks, well fitted as (asymmetric) Lorentz oscillators (dashed lines in Figure~\ref{fig2}). We attribute them to infrared-active lattice vibrations (phonons). The individual contributions of these phonons are neither displayed in Figure~\ref{fig2} nor are they discussed in this paper, in which the focus is on electronic excitations.  

Figure~\ref{fig8} compares the infrared conductivity spectra of K$_{3}$\! p-terphenyl (red line) and a bare p-terphenyl sample (black line) at {300}{K}. The latter exhibits an insulator-like response with a vanishing electronic conductivity, in good agreement with the reported band gap of about 3.5~eV~\cite{Puschnig_02}). In return, this confirms that the electronic conductivity of the K$_{3}$\! p-terphenyl sample originates from the electrons that are transferred from the K atoms to the $\pi$ orbitals of the p-terphenyl molecules.\\ 
The total spectral weight of the Drude and Lorentz bands allows us to estimate the number of electrons participating in these transitions. Assuming a free-electron mass, $m_i=m_e$, we determine the number of carriers per p-terphenyl molecule, $N_{\text{eff}}$, through the equation 

\begin{align}\label{eq3}
N_{\text{eff}} = \dfrac{N}{N_{0}} = \dfrac{1}{N_{0}} \dfrac{2m_{e}}{\pi e^{2}}\cdot \text{SW}_\text{tot},
\end{align}

where $N_0$ is the volume density of p-terphenyl molecules (about $32.5\times10^{20}$~cm$^{-3}$).
We obtain $N_{\text{eff}}$ $\approx$ 0.09. Considering a strong anisotropy of the intra- and inter-molecular electronic excitations and assuming a powder average of randomly oriented grains that are smaller than the wavelength, this value corresponds to about 10~$\%$ of the one expected for nominally three electrons per molecule. This discrepancy can be explained (at least partially) by the low volume density of the sample which has been pressed from powder, by an electron mass that exceeds $m_{e}$, and by the lack of an effective medium description that accounts for the shape of the crystallites and their particular electronic anisotropy (see e.g.~\cite{sihvola1999electromagnetic}). However, this discrepancy might also be an indication that the number of $\pi$ electrons per molecule is somewhat less than the nominal 3 electrons.

The low-frequency part of the electronic conductivity spectrum contains a weak very Drude peak at the origin and the similarly weak Lorentzian mode $L_1$. A linear extrapolation at {300}{K} towards zero frequency, or likewise the fitted parameters $\omega_{p0}$ and $\Gamma_0$, yield a dc conductivity of about 4~$\Omega^{-1}$cm$^{-1}$ that matches reasonably well the dc-resistivity measured on a pellet from the same growth batch. The spectral weight of the Drude peak is about two orders of magnitude lower than that of the L$_{3}$ band (see Table~\ref{TableLat}). Such a small free carrier spectral weight is not uncommon for low-dimensional organic conductors for which the inter-molecular hopping parameter is typically much smaller than the intra-molecular ones~\cite{Jerome_04, Dressel_12}. We will come back to this point in Section \ref{sec:assignment}.

\section{Assignment of the main transitions}\label{sec:assignment}

H\"uckel theory is quite useful for discussing qualitatively the optical spectra of organic molecular solids. This has been demonstrated in the case of C$_{60}$, where the discrete molecular energy levels are broadened into narrow bands~\cite{Gunnarsson_97} in such a way that the level structure of the molecule remains visible in the optical absorption spectrum of the molecular solid~\cite{Degiorgi_98}.
In H\"uckel's theory of  conjugated hydrocarbons $\pi$ electrons are treated explicitly, in tight-binding approximation, while $\sigma$ electrons are taken into account through the elastic energy of bond-stretching \cite{Huckel_31, Huckel_32, Salem_66}. The hopping amplitudes are supposed to vary linearly with respect to changes in bond lengths around an average value $t_0$, with a slope $\alpha$. The elastic energy involved in bond stretching depends on a force constant $K$. The theory is worked out in detail for a p-terphenyl molecule in Appendix~\ref{sec:huckel}, using parameters $t_0,\alpha, K$ extracted from electronic and vibrational properties of other molecules (such as benzene). Therefore we do not use any adjustable parameters. We should however keep in mind that the parametrization has been made for neutral molecules. Some modifications due to charge transfer have been discussed in the case of intercalated graphite \cite{Pietronero_81}, but we leave the parameters $t_0,\alpha,K$ unchanged and limit ourselves to changes in the bond order (see Appendix~\ref{sec:huckel}).

\begin{figure}[h] 
	\includegraphics[width=0.95\linewidth]{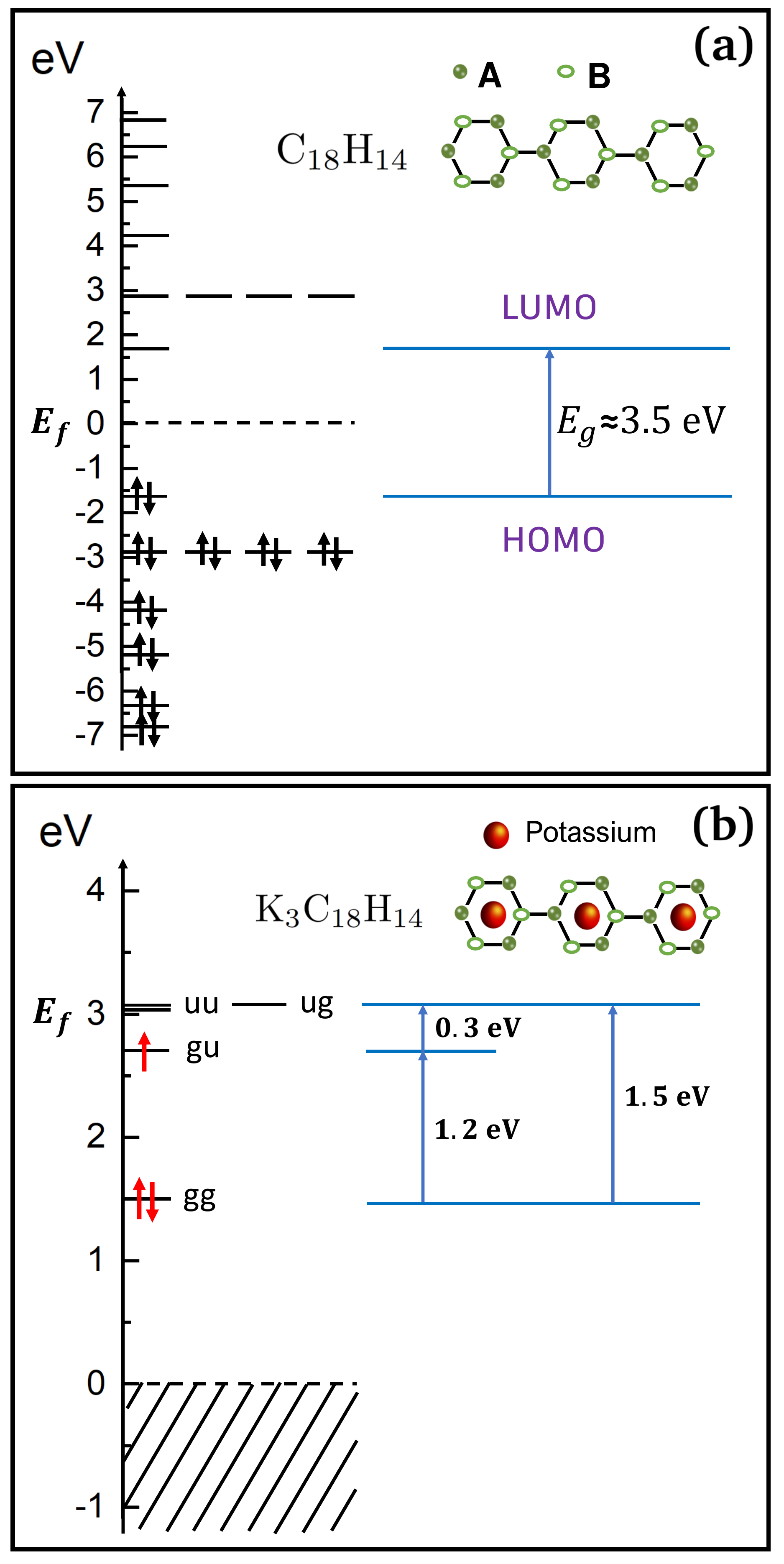}
	\caption{Energy levels and lowest optical transitions according to the H\"uckel model. (a) Hypothetical case of pristine p-terphenyl with equal bond lengths. (b) K$_{3}$\,\!p-terphenyl for optimized bond lengths.}
	\label{fig:huckel}
\end{figure}

\B{Figure~\ref{fig:huckel}(a)} shows the energy levels of a neutral p-terphenyl molecule for the case of equal bond lengths. The spectrum has particle-hole symmetry, because the system is bi-partite in tight-binding approximation, i.e., the sites can be subdivided into two sublattices $\mathcal{A}$ and $\mathcal{B}$ in such a way that hopping occurs only between the two sublattices, as illustrated in Figure \ref{fig:huckel}(a). The energy difference between the lowest unoccupied orbital (LUMO) and the highest occupied orbital (HOMO) is about 3.5 eV, in good agreement with the energy gap reported for pristine p-terphenyl. In contrast to C$_{60}$, where the LUMO is threefold degenerate (even sixfold for equal bond lengths~\cite{Wang_93}), the LUMO (as well as the HOMO) is non-degenerate in the case of p-terphenyl. The next level (``Lumo+1'') is four-fold degenerate, but it is split (of the order of 20 meV) into two non-degenerate and one doubly degenerate level if the bond lengths are optimized. Much larger ``polaronic'' shifts (of the order of 0.3 eV) occur if we add electrons, as illustrated in Figure~\ref{fig:huckel}(b) for the case of K$_3$ p-terphenyl. We classify the orbitals according to their parity with respect to reflections by the long and short axes. Thus $gg$ means even (``gerade'') with respect to both reflections, $gu$ means even with respect to the first and odd (``ungerade'') with respect to the second reflection, and so on. Optical transitions are symmetry-allowed for $gg\rightarrow gu$, $gg\rightarrow ug$ and $gu\rightarrow uu$, with transition energies as indicated in Figure \ref{fig:huckel}(b).

The optical conductivity can also be calculated in the framework of H\"uckel theory, as detailed in Appendix~\ref{sec:huckel}. The symmetry of molecular orbitals implies selection rules, as expected, namely the transition $gg\rightarrow gu$ occurs for light polarized parallel to the long axes, while the transitions $gg\rightarrow ug$ and $gu\rightarrow uu$ require the polarization to be parallel to the short axis. Since the experiments were performed on polycrystalline samples we have calculated the powder averaged response. The results are shown in Table \ref{tab:conductivity}. 

\begin{table}
\centering
$
\begin{array}{ccc}
\mbox{Transition}&\mbox{Transition energy}&\mbox{Spectral weight}\\
\hline
gg\rightarrow gu&1.232&1.396\times 10^6\\
gg\rightarrow ug&1.522&0.340\times 10^6\\
gu\rightarrow uu^{(1)}&0.287&8.369\times 10^6\\
gu\rightarrow uu^{(2)}&0.290&9.505\times 10^6
\end{array}
$
\caption{Low-energy intra-molecular transitions of K$_3$\! p-terphenyl according to H\"uckel theory. The transition energies are given in units of eV, the oscillator strengths in units of $\Omega^{-1}$cm$^{-2}$.}
\label{tab:conductivity}
\end{table}

The predominant features of the electronic response of the K$_3$\! p-terphenyl sample are the Lorentzian modes $L_{3}$ and $L_{4}$. We show now that they agree reasonably well with the transitions predicted by H\"uckel theory for a tri-anionic p-terphenyl molecule. The $L_{3}$ band with an eigenfrequency of $\omega_3=4000$ cm$^{-1}$ (0.5 eV) can be roughly accounted for by the dipole-allowed $gu\rightarrow uu$ transitions around 0.3 eV (Figure \ref{fig:huckel} and Table \ref{tab:conductivity}). Likewise, the $L_{4}$ band with $\omega_4=12000$ cm$^{-1}$ (1.5 eV) can be well described by the $gg\rightarrow gu$ and $gg\rightarrow ug$ transitions at 1.2 eV and 1.5 eV, respectively.

The relatively poor agreement between theory and experiment in the case of the $L_3$ transition (0.3 versus 0.5 eV) can be understood as follow. 
Electron-electron correlations are known to produce non-negligible effects both for the ground state and for the excited states of hydrocarbons~\cite{Baeriswyl_92}. H\"uckel theory also predicts a ``bipolaronic'' path for doping, as shown in Appendix~\ref{sec:huckel}. If Coulomb repulsion among $\pi$ electrons is taken explicitly into account, this scenario is not expected to be energetically favored.

The comparison of the spectral weights obtained from theory and experiment is more tricky. Clearly, the theoretical values of Table \ref{tab:conductivity} are substantially larger than the experimental values of Table \ref{TableLat}. The discrepancy in the relative spectral weights is even more striking. While the experimental values for the $L_{3}$ mode are about 1.5 times larger than those for the $L_{4}$ mode, the theoretical numbers for the $gu\rightarrow uu$ transitions are an order of magnitude larger than those for the $gg\rightarrow gu$ and $gg\rightarrow ug$ transitions. The main difference between the two types of transitions is that the former depend strongly on the occupation of the level at 2.7 eV, while the latter are less sensitive. The agreement of the relative intensities thus could be greatly improved by postulating that the actual charge per molecule is less than 3, either due to an only partial electron transfer from the K ions to the p-terphenyl molecules or a K concentration that is lower than the nominal one. Evidence for a lower effective electron doping has been also found from the experiments in Ref.~\cite{pinto2020potassium} and the calculations in Ref.~\cite{Guijarro_20}. 

 The corresponding inter-molecular hopping is expected to be much weaker than the intra-molecular one and will therefore mainly contribute to the broadening of the intra-molecular levels as shown in Figure 4 (in addition to defects and a possible variation of the local K-content). Moreover, it accounts for the weak Drude peak that arises from the inter-molecular excitations of the single electron in the $gu$-level at 2.7eV and likely also for the weak L$_1$ mode at 220 cm$^{-1}$. The latter could be explained by a variation of the local K-content and the subsequent transfer of electrons to the p-terphenyl molecules, which would give rise to small changes of the intra-molecular energy levels. 
\begin{figure*} [t]
	\includegraphics[width=0.95\linewidth]{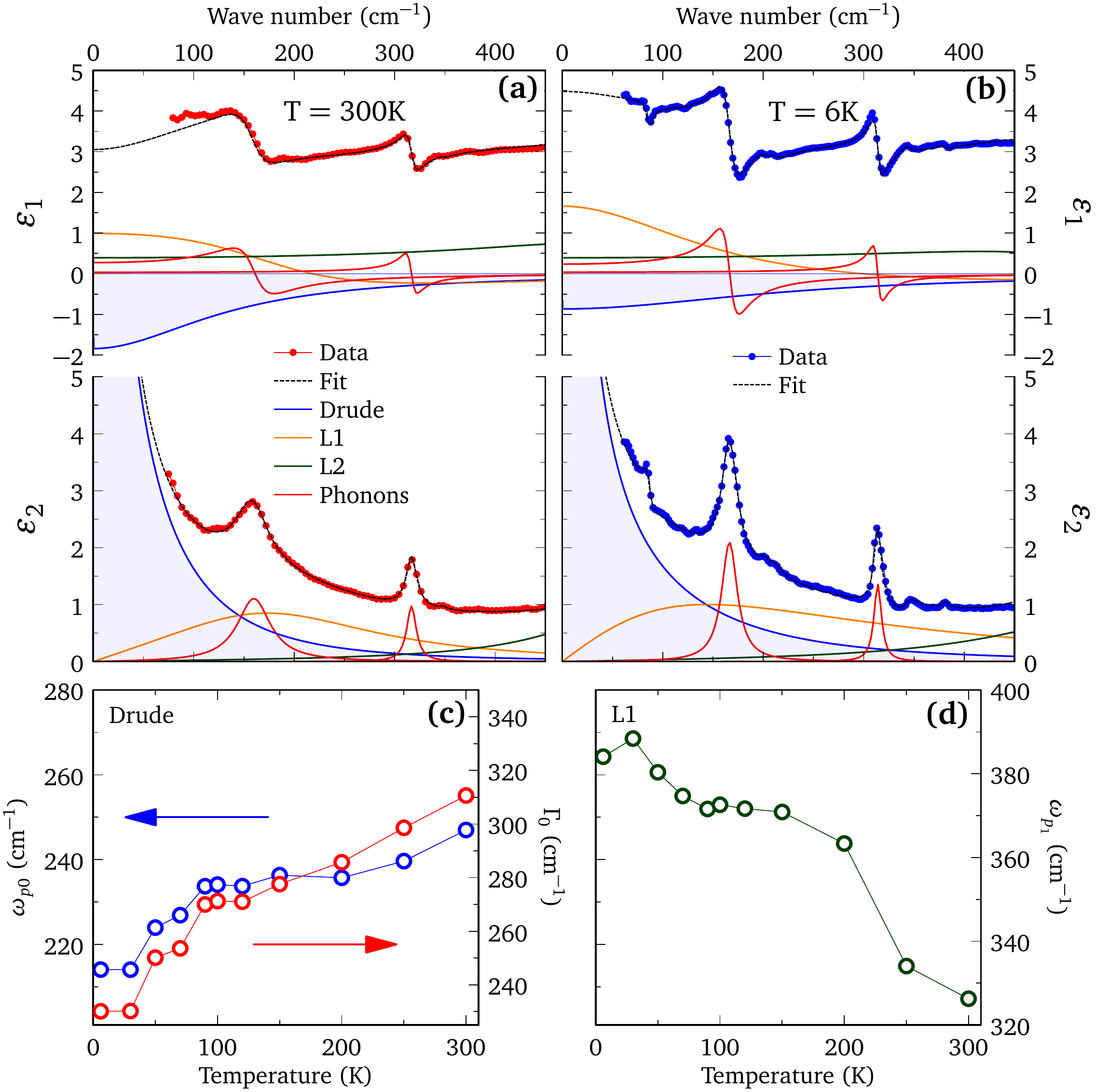}
	\caption{(a) Spectra at {300}{K} of the real and imaginary parts of the dielectric function (symbols). The dashed lines show the total fit function and the solid lines the individual contributions of the Drude response (blue), the low-energy Lorentz oscillators (orange and green) and of two phonon modes (red). (b) Corresponding spectra (symbols) and fits (lines) measured at {6}{K}. (c) Temperature dependence of the obtained fit parameters for the scattering rate, $\Gamma_{0}$, and the plasma frequency, $\omega_{p0}$, of the Drude peak and (d) of the plasma frequency of the low-energy Lorentz oscillator, $\omega_{p1}$.}
	\label{fig3}
\end{figure*}

\section{Temperature dependence of the low-frequency response}\label{sec:low}

\begin{figure} [h]
	\includegraphics[width=0.95\linewidth]{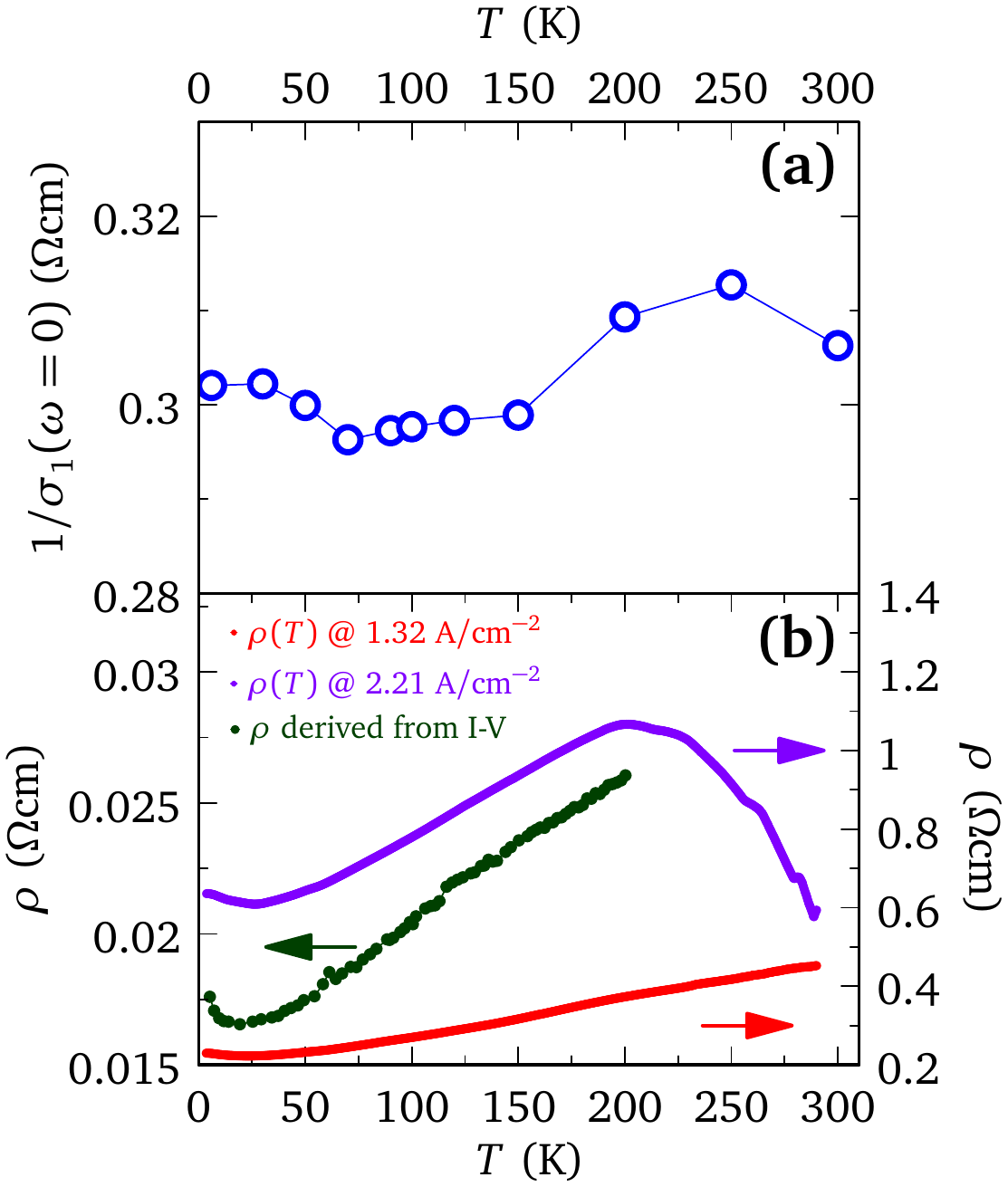}
	\caption{(a) $T$ dependence of the inverse of the dc conductivity of a K$_{3}$\,\!p-terphenyl sample, derived from a fit of the IR spectra (extrapolated at $\omega$ = 0). (b) Electrical resistivity of a twin sample measured with sourcing current pulses of $J= 1.32$  A/cm$^2$ (red points) and $J= 2.21$ A/cm$^2$ (violet points). Also shown are data derived from the slope of the low current range branch of the {\it{I-V}} curve (dark green points).}
	\label{fig5}
\end{figure}

Next, we discuss the temperature dependence of the weak Drude peak and of the low-energy Lorentzian $L_{1}$. Figures~\ref{fig3}(a) to ~\ref{fig3}(b) show the measured spectra (symbols) of the real and imaginary parts of the dielectric function,~$\varepsilon_{1}$ and~$\varepsilon_{2}$, up to 450 cm$^{-1}$ at the highest and the lowest temperatures of {300}{K} and {6}{K}, respectively. The black dashed line shows the best fit to the complex dielectric function in the range below 450 cm$^{-1}$ using a Drude-Lorentz model similar to that of equation (1). The colored lines show the individual contributions of the Drude peak (blue), the $L_{1}$ (orange) and $L_2$ peaks (dark green), and of the low-energy IR phonons (red). To minimize the number of fit parameters, we fixed the position and width of the $L_1$ peak as well as the position and weight of $L_2$ peak at the values obtained at 300 K.
The obtained temperature dependence of the plasma frequency and width of the Drude peak and the plasma frequency of the $L_{1}$ peak are displayed in Figures~\ref{fig3}(c) and \ref{fig3}(d). For the Drude response both the plasma frequency~$\omega_{p0}$ and the scattering rate~$\Gamma_{0}$ decrease substantially as the temperature is reduced below 300 K. Moreover, for both Drude parameters the temperature dependent decrease appears to be anomalously enhanced below about {90}{K}. The plasma frequency of the $L_{1}$ peak is also temperature-dependent and increases towards low temperature. Notably, the spectral weight loss of the Drude peak between {90}{K} and {6}{K} of about 228 $\Omega^{-1}$cm$^{-2}$ matches rather well the corresponding spectral weight gain of the $L_{1}$ peak which amounts to about 244 $\Omega^{-1}$cm$^{-2}$. Overall, the fitting of the low-energy response provides evidence for a low concentration of mobile carriers that tend to get localized as the temperature is decreased. Why this localization is anomalously enhanced below about {90}{K} remains to be understood.  

We now turn to the comparison of the charge transport, as determined from the infrared spectra and the resistivity data. 
Figure~\ref{fig5}(a) shows the temperature dependence of the inverse dc conductivity, derived from the fit parameters $\omega_{0D}$ and $\Gamma_{D}$ of the Drude peak.

The temperature dependent electrical resistivity, $\rho(T)$, has been detected in a twin K$_{3}$\,\!p-terphenyl sample and is displayed in Figure~\ref{fig5}b. The values have been obtained at two different, fixed values of the current density, $J$, as well as from the {\it{I-V}} curves, taking into account the slope of the branch of the {\it{I-V}} curve measured in the low current range. The resistivity behavior is non-ohmic, with features that are correlated with the current density. 
Generally, $\rho(T)$ decreases as $T$ is lowered from {300}{K} (except at $J= 2.21$ A/cm$^2$ where  $\rho(T)$ first increase to about {200}{K} before it decreases), saturating around {30}{K} and then upturning below {30}{K}. The low tempeature upturn of $\rho(T)$ becomes more pronounced as the sourced current density is increased. Nevertheless, it is also evident in the curve derived from the {\it{I-V}} characteristics (Figure~\ref{fig5}b). The largest changes in the resistivity have been detected at the highest current density ($J= 2.21$ A/cm$^2$), where $\rho(T)$ exhibits a broad, bell shaped maximum centered around {200}{K} that closely resembles the behavior shown in Figure~\ref{fig5}a. A bell shaped resistivity as a function of temperature has been also obtained in a previous study of some of the authors on KPT electrical transport properties in Ref.~\cite{pinto2020potassium}. The values of $\rho(T)$ derived from the IR spectra are of the same order of magnitude as those of the electrical measurements at high currents of $J = 1.32$ A/cm$^2$ and  $J = 2.21$ A/cm$^2$ and about one order of magnitude higher than those derived from the {\it{I-V}} curves measured at fixed $T$ (Figure~\ref{fig5}a \& \ref{fig5}b).\\
These observations seem compatible with the scenario of a spatially inhomogenous conductivity with conducting patches that are embedded in a matrix of materials that is less conducting. Here, the transport measurements at low current probe primarily the conducting filaments whereas, at a higher current, they are more representative of the averaged conductivity (similar to the infrared data).
The temperature-dependent dc resistivity deduced from the IR spectra, exhibits some features that are also detected in the electrical transport data at high $J$ value, such as the broad maximum around {200}{K} and the low-$T$ upturn.\\ 
It's worthwhile noting that, the low-$T$ upturn of $\rho(T)$ detected by IR below {75}{K}, occurs
at a somewhat lower $T$ than that of the suppression of the plasma frequency and the scattering rate in Figure 4c. This is due to a partial compensation of the effects of the reduced plasma frequency and scattering rate on the derived value of the dc resistivity. Irrespective of these minor differences, the low-$T$ upturn of $\rho(T)$ appears to be a feature that is common to the transport and infrared data and thus intrinsic to the charge dynamics. 
In particular, the suppression of the spectral weight of the Drude response below {90}{K} and the related spectral weight gain of the L$_{1}$-band around 220~cm$^{-1}$ in the infrared spectra is indicative of weak localization of some of the charge carriers. Notably, it is inconsistent with the formation of a macroscopic superconducting state for which the missing spectral weight would be red-shifted and give rise to a delta-function at the origin, that accounts for the loss-free response of the superconducting condensate. 
The infrared spectra are therefore in contradiction with the scenario that a macroscopic superconducting state develops below {90}{K} in our K$_{3}$\,\!p-terphenyl sample. On the other hand, they do not exclude the possibility of a filamentary superconducting state in a small volume fraction of the sample, well below the threshold for percolation. In this context, we note that low dimensional structures in the form of interconnected filaments have indeed been proposed as a source of amplification of superconductivity by shape resonance effects in Ref. \cite{Bianconi2017}. The plasmonic effects related to a spatial confinement of the SC condensate could indeed give rise to a plasmonic mode that is centered at finite frequency~\cite{Wang_12, sihvola1999electromagnetic}. The formation of such a plasmonic feature thus could possibly explain the blue-shift of some of the spectral weight of the Drude peak to the L$_{1}$-mode that sets in below {90}{K}. 

\section{Discussion and outlook}\label{sec:discussion}
In summary, with infrared ellipsometry we have studied the electronic response of a polycrystalline sample of K$_{3}$\,\!p-terphenyl. For this purpose we have mounted the sample, that decompose rapidly under humid ambient conditions, inside a glovebox under dry and inert Argon gas atmosphere inside a cell with windows for optical measurements that can be sealed against the ambient. 
The measured spectra of the electronic conductivity are governed by two pronounced Lorentzian bands with maxima around 4\,000\,cm$^{-1}$  and  12\,000\,cm$^{-1}$. Based on the comparison with calculations using a H\"uckel model, these bands have been assigned to intra-molecular excitations of the $\pi$ electrons of the anionic p-terphenyl molecules. Some discrepancies between experiment and theory, in particular, with respect to the relative spectral weights of these bands, can be explained if the molecules contain less than the nominal three $\pi$ electrons (but still more than two), in agreement with an effective number of K atoms per p-terphenyl molecule lower than 3 detected experimentally~\cite{pinto2020potassium}. 
The corresponding inter-molecular excitations give rise to a very weak Drude peak at the origin and most likely also to a weak Lorentzian band around 220 cm$^{-1}$. The dc resistivity (inverse conductivity) derived from the Drude response is consistent with the behavior of the dc resistivity from electrical transport measured at high current density. As the temperature is reduced below {300}{K} the plasma frequency and the width of this Drude peak exhibit a gradual decrease that is anomalously enhanced below about {90}{K}. The related missing spectral weight of the Drude peak at low temperature is blue-shifted toward higher energy where it increases the spectral weight of the band at 220 cm$^{-1}$. Such a blue-shift of low-energy spectral weight is not uncommon for such low-dimensional organic conductors for which the carriers tend to get weakly localized as temperature decreases. It certainly contradicts the behavior expected for a bulk-like superconductor for which the missing spectral weight of the Drude peak should be red-shifted to a delta-function at the origin that accounts for the loss-free response of the superconducting condensate. Nevertheless, the infrared data are not necessarily incompatible with a spatially inhomogenous superconducting state with a volume fraction well below the percolation limit for which the confinement of the superconducting condensate might result in a plasmonic resonance at finite frequency.


\begin{acknowledgments}
Work at University of Fribourg was supported by the Schweizerische Nationalfonds (SNF) by Grant No. 200020-172611. We thank M. Andrey from the mechanical workshop at UniFr for the design and building of the optical cell. 
\end{acknowledgments}

\appendix
\section{H\"uckel theory for K$_x$\! p-terphenyl}\label{sec:huckel}
 In this appendix we use H\"uckel theory to calculate the equilibrium structure, the electronic energy levels and the optical absorbtion for a p-terphenyl molecule. The bond lengths $r_{ij}$ between neighboring sites $i$, $j$ are assumed to remain close to an average value $r_0$. We write
\begin{align}
r_{ij}=r_0+u_{ij}
\end{align}
and expand the hopping amplitudes as
\begin{align}
t(r_{ij})=t_0-\alpha u_{ij}\, ,
\end{align}
 where $t_0$ is the value at $r_0$. The Hamiltonian reads
 \begin{align}\label{eq:hamiltonian}
 H=&\frac{K}{2}\sum_{\langle i,j\rangle} u_{ij}^2-\sum_{\langle i,j\rangle\sigma}(t_0-\alpha u_{ij})
 \big(c_{i\sigma}^\dag c_{j\sigma}^{\phantom{}}+c_{j\sigma}^\dag c_{i\sigma}^{\phantom{}}\big) \nonumber\\
 &-2\alpha p_0\sum_{\langle i,j\rangle} u_{ij}\, ,
 \end{align}
where the operators $c_{i\sigma}^\dag$ ($c_{i\sigma}^{\phantom{}}$) create (annihilate) $\pi$ electrons at site $i$ with spin $\sigma$. The first term is the elastic energy, the second term describes hopping of $\pi$ electrons between neighboring sites, and the third term fixes the average bond length to $r_0$. The quantity $p_0$ is the average expectation value of the bond order operator
\begin{align}
p_{ij}=\frac{1}{2}\sum_{\sigma}
 \big(c_{i\sigma}^\dag c_{j\sigma}^{\phantom{}}+c_{j\sigma}^\dag c_{i\sigma}^{\phantom{}}\big)\, .
\end{align}
The derivatives of the ground state energy 
\begin{align}
\frac{\partial}{\partial u_{ij}}\langle H\rangle=Ku_{ij}+2\alpha (\langle p_{ij}\rangle-p_0)
\end{align}
vanish at the energy minimum, where one obtains the linear relation
\begin{align}\label{eq:bondorderbondlength}
u_{ij}=-\frac{2\alpha}{K}(\langle p_{ij}\rangle-p_0)\, .
\end{align}
For systems such as benzene $\langle p_{ij}\rangle$ is the same for all bonds and equal to $p_0$, therefore $u_{ij}$ vanishes on all bonds, as it should. For terphenyl $\langle p_{ij}\rangle$ is not constant and therefore the bond lengths are not all equal. Eq. (\ref{eq:bondorderbondlength}) can also be used to relate two different systems with similar bond orders, such as benzene and graphene. In fact, an essentially linear function is obtained if we plot the average bond orders of different systems versus their average bond lengths \cite{Baeriswyl_85}. From the slope of these data one can deduce a value $\alpha/K\approx 0.1$ \AA.
\begin{figure}
\centering
\vspace{0.5cm}
\includegraphics[width=8cm]{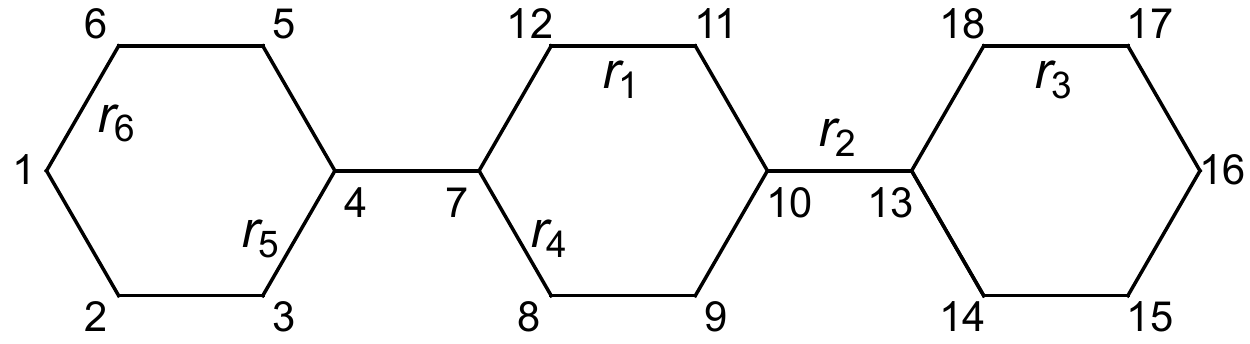}
\caption{Structure of $p$-terphenyl and numbering of $\pi$ orbitals and bond lengths.}
\label{fig:structure}
\end{figure}

We assume the molecule to be planar and symmetric with respect to reflections 
$\sigma_x$ and $\sigma_y$ by the long and short axes, respectively. It follows that there can only be 6 different bond lengths $r_i$, $i=1,...,6$, as illustrated in Figure \ref{fig:structure}. We now choose a basis of states which are symmetric or antisymmetric with respect to the two reflection operators, in terms of superpositions of (orthogonalized) $\pi$ orbitals $\vert i\rangle$, $i=1,...,18$.
\subsection{Symmetry-adapted basis}
There are six linearly independent states which are even with respect to both $\sigma_x$ and $\sigma_y$ and we choose them as ($g$ for ``gerade'')
\begin{align}\label{eq:gg}
\vert gg\rangle_1&=\frac{1}{2}( \vert 2 \rangle+ \vert 6\rangle+ \vert 15\rangle+ \vert 17\rangle)\nonumber\\
\vert gg\rangle_2&=\frac{1}{2}( \vert 3 \rangle+ \vert 5\rangle+ \vert 14\rangle+ \vert 18\rangle)\nonumber\\
\vert gg\rangle_3&=\frac{1}{2}( \vert 8 \rangle+ \vert 9\rangle+ \vert 11\rangle+ \vert 12\rangle)\nonumber\\
\vert gg\rangle_4&=\frac{1}{\sqrt{2}}(\vert 1\rangle+\vert 16\rangle)\nonumber\\
\vert gg\rangle_5&=\frac{1}{\sqrt{2}}(\vert 4\rangle+\vert 13\rangle)\nonumber\\
\vert gg\rangle_6&=\frac{1}{\sqrt{2}}(\vert 7\rangle+\vert 10\rangle)
\end{align}
There are also six different states which are even with respect to $\sigma_x$ and odd with respect to $\sigma_y$, namely ($u$ for ``ungerade'')
\begin{align}\label{eq:gg}
\vert gu\rangle_1&=\frac{1}{2}( \vert 2 \rangle+ \vert 6\rangle- \vert 15\rangle- \vert 17\rangle)\nonumber\\
\vert gu\rangle_2&=\frac{1}{2}( \vert 3 \rangle+ \vert 5\rangle- \vert 14\rangle- \vert 18\rangle)\nonumber\\
\vert gu\rangle_3&=\frac{1}{2}( \vert 8 \rangle-\vert 9\rangle- \vert 11\rangle+ \vert 12\rangle)\nonumber\\
\vert gu\rangle_4&=\frac{1}{\sqrt{2}}(\vert 1\rangle-\vert 16\rangle)\nonumber\\
\vert gu\rangle_5&=\frac{1}{\sqrt{2}}(\vert 4\rangle-\vert 13\rangle)\nonumber\\
\vert gu\rangle_6&=\frac{1}{\sqrt{2}}(\vert 7\rangle-\vert 10\rangle)
\end{align}
Three states are odd with respect to $\sigma_x$ and even with respect to $\sigma_y$,
\begin{align}\label{eq:ug}
\vert ug\rangle_1&=\frac{1}{2}( \vert 2 \rangle- \vert 6\rangle+ \vert 15\rangle- \vert 17\rangle)\nonumber\\
\vert ug\rangle_2&=\frac{1}{2}( \vert 3 \rangle- \vert 5\rangle+ \vert 14\rangle- \vert 18\rangle)\nonumber\\
\vert ug\rangle_3&=\frac{1}{2}( \vert 8 \rangle+ \vert 9\rangle- \vert 11\rangle- \vert 12\rangle)
\end{align}
Finally, three states are odd with respect to both reflections,
\begin{align}\label{eq:uu}
\vert uu\rangle_{1}&=\frac{1}{2}( \vert 2 \rangle- \vert 6\rangle- \vert 15\rangle+\vert 17\rangle)\nonumber\\
\vert uu\rangle_2&=\frac{1}{2}( \vert 3 \rangle- \vert 5\rangle- \vert 14\rangle+ \vert 18\rangle)\nonumber\\
\vert uu\rangle_3&=\frac{1}{2}( \vert 8 \rangle- \vert 9\rangle+ \vert 11\rangle- \vert 12\rangle)
\end{align}
\subsection{Energy levels and equilibrium bond lengths}
To find the ground state of the Hamiltonian (\ref{eq:hamiltonian}) we have to diagonalize the electronic part for hopping parameters $t_i:=t(r_i)$, where the bond lengths $r_i$ can have six different values. In the basis constructed above the Hamiltonian is block-diagonal with two $6\times 6$ blocks
\begin{align}
\left(\begin{array}{cccccc}
0&-t_3&0&-\sqrt{2}t_6&0&0\\
-t_3&0&0&0&-\sqrt{2}t_5&0\\
0&0&-st_1&0&0&-\sqrt{2}t_4\\
-\sqrt{2}t_6&0&0&0&0&0\\
0&-\sqrt{2}t_5&0&0&0&-t_2\\
0&0&-\sqrt{2}t_4&0&-t_2&0
\end{array}\right)
\end{align}
for the $gg$ and $gu$ sectors (with $s=+1$ in the $gg$ sector and $-1$ in the $gu$ sector), and two $3\times 3$ blocks 
\begin{align}
\left(\begin{array}{ccc}
0&-t_3&0\\
-t_3&0&0\\
0&0&-s t_1\\
\end{array}\right)
\end{align}
for the other two symmetries (with $s=+1$ in the $ug$ sector and $-1$ in the $uu$ sector). The $3\times 3$ matrices are easily diagonalized. There are two non-degenerate levels, one at $-t_1$ with $gu$ symmetry and one at $+t_1$ with $uu$ symmetry, and two doubly degenerate levels at $\pm t_3$. 
 
We discuss first the (hypothetical) case of equal bond lengths, i.e., $t_i=t_0$, $i=1,...,6$. 
The eigenvalues for the $ug$ and $uu$ sectors are now at $\pm t_0$. Two eigenvectors at $+t_0$ have $uu$ symmetry, one $ug$ symmetry, while two eigenvectors at $-t_0$ have $ug$ symmetry, one $uu$ symmetry. The eigenvalue equations for the $gg$ and $gu$ sectors read
\begin{align}
(E+st_0)(E^5-8t_0^2E^3+2st_0^3E^2+15t_0^4E-8st_0^5)=0.
\end{align}
Therefore we obtain an additional eigenvector at $+t_0$, with $gu$ symmetry, and one at $-t_0$, with 
$gg$ symmetry. It follows that the levels at $\pm t_0$ are four-fold degenerate, as illustrated in Figure
\ref{fig:huckel}. Only one positive eigenvalue, of $gg$ symmetry, is below $t_0$, at 0.59264 $t_0$. 

To determine the equilibrium bond lengths we use
parameter values $t_0=2.9$ eV, $K=46$ eV\AA$^{-2}$, $\alpha=4.5$ eV\AA$^{-1}$, in line with Kakitani's analysis of vibrational and electronic properties of small organic molecules \cite{Kakitani_74}. The values of $p_0$ are determined for equal bond lengths $r_{ij}=r_0$ and $r_0$ is related to $p_0$ as $r_0=(1.525-0.2 p_0)$\AA, in agreement with bond orders and bond lengths of benzene and graphene ($p_0=\frac{2}{3}$, $r_0=1.39$\AA\, for benzene and $p_0=0.525$, $r_0=1.42$\AA\, for graphene).
The bond lengths obtained by minimizing the total energy for different doping levels are given in Table \ref{tab:bondlengths_H}. The numbers agree surprisingly well with results of DFT calculations (Table \ref{tab:bondlengths_S}).

\begin{table}[H]
\centering
$
\begin{array}{ccccccccc}
x&r_1&r_2&r_3&r_4&r_5&r_6&p_0&E_0\\
\hline
0&1.387&1.454&1.390&1.402&1.400&1.393&0.619&-71.86\\
1&1.376&1.434&1.385&1.419&1.413&1.399&0.605&-69.81\\
2&1.365&1.414&1.380&1.437&1.425&1.404&0.590&-67.89\\
3&1.373&1.415&1.373&1.435&1.440&1.419&0.565&-64.59\\
4&1.381&1.417&1.365&1.432&1.456&1.433&0.540&-61.41
\end{array}
$
\caption{Equilibrium bond lengths (\AA), average bond orders $p_0$ and ground-state energies $E_0$ (eV) for different doping levels, according to H\"uckel theory.}
\label{tab:bondlengths_H}
\end{table}

\begin{table}[H]
\centering
$
\begin{array}{ccccccc}
x&r_1&r_2&r_3&r_4&r_5&r_6\\
\hline
0&1.390&1.484&1.392&1.402&1.403&1.394\\
1&1.378&1.450&1.387&1.427&1.427&1.402\\
2&1.372&1.418&1.382&1.447&1.451&1.414
\end{array}
$
\caption{Equilibrium bond lengths (\AA) for different doping levels, according to \cite{Sakamoto_08}.}
\label{tab:bondlengths_S}
\end{table}

The energy levels between 0 and 3.2 eV (which are relevant for the IR absorption) are listed in Table \ref{tab:levels_H}. For $x=0$ all these levels are unoccupied. For $x=1$ and $x=2$ the $gg$ level is, respectively, singly and doubly occupied. Correspondingly, the energy level is lowered substantially. A similar ``polaronic'' effect is seen if the $gu$ level is occupied, i.e., from $x=2$ to $x=3$. For $x=4$ the $gg$ and $gu$ levels are both doubly occupied.
\begin{table}[H]
\centering
$
\begin{array}{cccccc}
&gg&gu&uu^{(1)}&ug&uu^{(2)}\\
\hline
r_i=r_0&1.719&2.900&2.900&2.900&2.900\\
x=0&1.783&2.910&2.945&2.945&2.956\\
x=1&1.658&2.865&2.967&2.967&3.006\\
x=2&1.538&2.826&2.987&2.987&3.056\\
x=3&1.501&2.733&3.020&3.023&3.023\\
x=4&1.464&2.641&2.984&3.058&3.058
\end{array}
$
\caption{Lowest positive energy eigenvalues (in eV for $t_0$=2.9\! eV) for the H\"uckel model of K$_x$ p-terphenyl. The numbers of the first row have been calculated for equal bond lengths, the others for optimized geometries.}
\label{tab:levels_H}
\end{table}
\subsection{Optical transitions}
We now use the H\"uckel model for treating optical transitions between molecular orbitals. In first quantization the electronic part of the Hamiltonian is
\begin{align}
H_e=-\sum_{i,j}t_{ij}\vert i\rangle\langle j\vert\, ,
\end{align}
where $t_{ij}$ is only finite if the sites $i,j$ are nearest neighbors. In the spirit of tight-binding theory we define the velocity operator as the time derivative of the position operator
\begin{align}
{\bf R}:=\sum_i{\bf R}_i\vert i\rangle\langle i\vert\, ,
\end{align}
where ${\bf R}_i$ is the position of the $i$th atom in the molecule, i.e., 
\begin{align}
{\bf v}=\frac{i}{\hbar}[H,{\bf R}]=\frac{i}{\hbar}\sum_{i,j}({\bf R}_i-{\bf R}_j)t_{ij}\vert i\rangle\langle j\vert\, .
\end{align}
If inter-molecular transitions are neglected, the conductivity is just that of a single molecule multiplied by the number $N_0$ of molecules per unit volume. Linear-response theory then implies \cite{Lax_58}
\begin{align}
\sigma_{1,\alpha\alpha}(\omega)=\frac{N_0\pi e^2}{\hbar\omega}
\sum_{\mu,\nu}\vert\langle\mu\vert v_\alpha\vert\nu\rangle\vert^2\delta\big(\omega-\omega_{\nu\mu}\big)\, ,
\end{align} 
where $\alpha=x,y$ is the polarization and $\hbar\omega_{\nu\mu}=E_\nu-E_\mu$ is the transition energy between occupied states $\vert\mu\rangle$ and unoccupied states $\vert\nu\rangle$
(the labels $\mu,\nu$ include spin).

This approach has been applied before for polyacetylene \cite{Baeriswyl_83} using a constant nearest-neighbor hopping amplitude $t_0$ in the velocity operator. Later it has been shown that the length-dependence of $t_{ij}$ produces correction terms, which however are small \cite{Gebhard_97}. They will be neglected here.

It is straigth forward to calculate the matrix elements between the different orbitals. We consider the levels of Table \ref{tab:levels_H} for $x=3$, for which the non-vanishing matrix elements are
\begin{align}
\langle gg\vert {\bf v}\vert gu\rangle&= -0.3951\, i\frac{tr_0}{\hbar}{\bf e}_x\, ,\\
\langle gg\vert {\bf v}\vert ug\rangle&= -0.1533\, i\frac{tr_0}{\hbar}{\bf e}_y\, ,\\
\langle gu\vert {\bf v}\vert uu\rangle^{(1)}&= -0.4670\, i\frac{tr_0}{\hbar}{\bf e}_y\, ,\\
\langle gu\vert {\bf v}\vert uu\rangle^{(2)}&= -0.5001\, i\frac{tr_0}{\hbar}{\bf e}_y\, ,
\end{align}
where ${\bf e}_x$ and ${\bf e}_y$ are unit vectors in the direction of the long and short axis, respectively. The polarization dependence reflects the symmetry of the molecular orbitals.

The spectral weight for the radiative transition $\mu\rightarrow\nu$ with $\alpha$-polarized light is given by

\begin{align}
\int_0^\infty
d\omega\, \sigma_{1,\alpha\alpha}^{(\mu\rightarrow\nu)}
=\frac{N_0\pi e^2}{\hbar\omega_{\nu\mu}}\vert\langle\mu\vert v_\alpha\vert\nu\rangle\vert^2\, .
\end{align}

To obtain the average values of Table \ref{tab:conductivity} one has to sum over the {polarization components and divide the result by 3. Moreover the spin gives a factor of 2 for the $gg\rightarrow ug$ transition. In the above analysis we have omitted the transition from the $gu$ level (at 2.7 to 3 eV) to the second $gg$ level (at 4eV) because its intensity is about 100 times smaller than the other transitions in the same energy range.

\subsection{Doping}
In the neutral state of the p-terphenyl molecule (18 $\pi$ electrons) all negative energy levels are filled. To describe how doping proceeds in a system of $N$ (uncoupled) molecules it is sufficient to consider $N=2$. A single additional electron occupies the lowest available level of one of the two molecules. This costs an energy $E_0(1)-E_0(0)$, where $E_0(x)$ is the ground state energy of a molecule for $x$ added electrons, as listed in Table \ref{tab:bondlengths_H}. One would naively expect that a second electron would be added to the other molecule, but another option is to add it to the same, with an energy cost $E_0(2)-E_0(1)$. The numbers of Table \ref{tab:bondlengths_H} show that the second choice requires less energy, explicitly
\begin{align}
E_0(2)-2E_0(1)+E_0(0)=-0.13\, \mbox{eV}.
\end{align}
We can also say that the state of two singly charged molecules is unstable with respect to a ``phase separation'' into a neutral and a doubly charged molecule. The same arguments show that two molecules with $x=3$ have a higher total energy than one with $x=4$ and one with $x=2$, because
 \begin{align}
E_0(4)-2E_0(3)+E_0(2)=-0.12\, \mbox{eV}.
\end{align}
Therefore according to H\"uckel's theory doping proceeds by adding electrons pairwise (by producing ``bipolarons'' instead of ``polarons'').


%

\end{document}